\documentclass[%
 reprint,
 amsmath,amssymb,
 aps,
prl,
floatfix,
]{revtex4-1}

\usepackage{graphicx}
\usepackage{dcolumn}
\usepackage{bm}
\usepackage{color}

\begin{document}

\title{Engineering of a Low-Entropy Quantum Simulator for Strongly Correlated Electrons Using SU($\mathcal{N}$)-Symmetric Cold Atom Mixtures}
\author{Daisuke Yamamoto$^{1}$}
\email{yamamoto.daisuke21@nihon-u.ac.jp}
\author{Katsuhiro Morita$^2$}
\email{katsuhiro.morita@rs.tus.ac.jp}
\affiliation{$^1$Department of Physics, College of Humanities and Sciences, Nihon University, Sakurajosui, Setagaya, Tokyo 156-8550, Japan}
\affiliation{$^2$Department of Physics, Faculty of Science and Technology, Tokyo University of Science, Chiba 278-8510, Japan}
\date{\today}
\begin{abstract}
An advanced cooling scheme, incorporating entropy engineering, is vital for isolated artificial quantum systems designed to emulate the low-temperature physics of strongly correlated electron systems (SCESs). This study theoretically demonstrates a cooling method employing multi-component Fermi gases with SU($\mathcal{N}$)-symmetric interactions, focusing on the case of $^{173}$Yb atoms in a two-dimensional optical lattice. Adiabatically introducing a nonuniform state-selective laser gives rise to two distinct subsystems: a central low-temperature region, exclusively composed of two specific spin components, acts as a quantum simulator for SCESs, while the surrounding $\mathcal{N}$-component mixture retains a significant portion of the entropy of the system. The SU($\mathcal{N}$)-symmetric interactions ensure that the total particle numbers for each component become good quantum numbers, creating a sharp boundary for the two-component region. The cooling efficiency is assessed through extensive finite-temperature Lanczos calculations. The results lay the foundation for quantum simulations of two-dimensional systems of Hubbard or Heisenberg type, offering crucial insights into intriguing low-temperature phenomena in condensed-matter physics.
\end{abstract}
\maketitle

Simulating a large-scale quantum-mechanical system poses a formidable computational challenge in nearly all areas of physics. Recent remarkable developments in experimental techniques have paved the way to directly simulate complex many-body physics in a quantum system by using an alternative controllable system realized on experimental platforms such as ultracold atomic and molecular gases~\cite{Bloch2005,Bloch2012,Gross2017,TAKAHASHI2022}, Rydberg atom arrays in optical tweezers~\cite{Nogrette2014,Bernien2017,Browaeys2020,Scholl2021,Ebadi2021}, trapped ions~\cite{Blatt2012,Monroe2021}, photonic systems~\cite{AspuruGuzik2012,Noh2016}, quantum dots~\cite{Hensgens2017}, and superconducting circuits~\cite{Houck2012}. The applications of such quantum simulations extend across a wide range of issues in diverse fields, including condensed-matter physics, atomic physics, quantum chemistry, high-energy physics, and cosmology~\cite{Cirac2012,Georgescu2014,Hempel2018,Altman2021,Bauer2023}. 

Engineering a low-temperature quantum system of Fermi particles with two internal states is of particular importance in the realm of quantum simulation studies~\cite{Zhang2000,Lewenstein2007,Hofstetter2018}. This is attributed to the fact that electrons, possessing a spin quantum number of 1/2, play a key role in solid-state physics. Particularly within strongly-correlated electron systems (SCESs), various phenomena bear both fundamental scientific importance and practical applications, such as Mott insulators, high-temperature superconductivity~\cite{Lee2006}, quantum magnetism~\cite{Auerbach1994}, geometric frustration~\cite{Moessner2006}, the Kondo effect~\cite{Kouwenhoven2001}, and more. A straightforward method to replicate these SCESs involves confining cold fermionic atoms, e.g., $^6$Li, with two different hyperfine states in an optical lattice potential~\cite{Jrdens2008,Schneider2008,Esslinger2010,Greif2013,Duarte2015,Hart2015,Greif2016,Parsons2016,Boll2016,Cheuk2016sc,Cheuk2016,Drewes2017,Mazurenko2017,Brown2017}. While this setup is advantageous for creating large-size lattice systems of the Hubbard type in any dimension, achieving low temperatures to observe highly quantum phenomena has posed a longstanding and significant challenge.  

Cold-atom systems, well isolated from the thermal environment, require precise entropy control for studying low-temperature physics. 
In the recent work of Mazurenko $et$ $al$.~\cite{Mazurenko2017}, a meticulously designed confinement potential was prepared using a digital micromirror device to divide the system into two subsystems: a central disk-shaped region comprising approximately 80 sites, each hosting nearly one atom, and a larger surrounding region with significantly lower density. The sparsely populated atoms in the latter subsystem form a metallic phase, serving as an entropy reservoir due to their high degree of mobility and effectively cooling down the central target region~\cite{Bernier2009}. This approach has successfully generated a low-entropy state of two-component fermions, exhibiting long-range antiferromangetic correlations in a two-dimensional optical lattice~\cite{Mazurenko2017}. However, achieving even lower temperatures, essential for studying phenomena like high-temperature superconductivity and quantum spin liquids~\cite{Savary2016}, requires an additional ingenious twist in conjunction with this entropy engineering method utilizing the motional degrees of freedom. 
 
In this Letter, we explore an entropy engineering scheme utilizing SU($\mathcal{N}$) atomic gases, aiming to use it for simulating two-dimensional quantum SCESs. Recent advancements in manipulating cold alkaline-earth-metal(-like) atoms, including $^{173}{\rm Yb}$ and $^{87}{\rm Sr}$~\cite{Fukuhara2007,Cazalilla2009,Gorshkov2010,DeSalvo2010,Tey2010,Taie2010,Cazalilla2014,Zhang2014,Hofrichter2016,Ozawa2018,Taie2022}, have spurred extensive investigations into quantum many-body systems with SU($\mathcal{N}$) symmetry, where $\mathcal{N} > 2$. This surge in research has led to predictions of exotic ground states for various lattice geometries and different values of the number of components $\mathcal{N}$~\cite{Tth2010,Bauer2012,Nataf2014,Corboz2011,Hermele2011,Romen2020}, scenarios not typically observed in electron systems limited to SU(2) or lower symmetry. Studies on the effects of external fields imposing a global population imbalance among spin components have also been conducted~\cite{Yamamoto2020,Motegi2023,Miyazaki2022}.

The SU($\mathcal{N}$) systems, especially those with a large $\mathcal{N}$, offer enhanced cooling efficiency, akin to the Pomeranchuk cooling mechanism~\cite{Richardson1997}. Here, we capitalize on this advantage and employ an SU($\mathcal{N}$) subsystem as an entropy reservoir, achieved through shaping a spin-dependent field potential. While the concept of the entropy engineering using spin degrees of freedom has been explored in the seminal work of Ref.~\onlinecite{ColomTatch2011}, it was limited to an exactly-solvable one-dimentional spin-3/2 chain. Quantum many-body systems on a two-dimensional lattice at finite temperatures, as considered here, are directly relevant to long-standing issues in SCESs. However, they pose a numerical challenge, especially when dealing with a large number of local states. Below, we perform extensive numerical computations using the finite-temperature Lanczos (FTL) method~\cite{Jakli1994,Prelovek2013} to demonstrate the efficiency of the entropy engineering scheme employing an SU($\mathcal{N}$) entropy reservoir.

We model an SU($\mathcal{N}$)-symmetric Fermi gas in an optical lattice by the following $\mathcal{N}$-component Hubbard Hamiltonian with spin-independent hoppings ($t$) and interactions ($U>0$)~\cite{Honerkamp2004-sj}:
\begin{eqnarray*}
\hat{\mathcal{H}}_{\rm Hub}&=&-t\sum_{\langle i,j\rangle;\sigma}\left(\hat{{c}}_{i,\sigma}^\dagger \hat{{c}}_{j,\sigma}+{\rm H.c.}\right) +U\sum_{\sigma<\sigma^\prime}\hat{n}_{i,\sigma}\hat{n}_{i,\sigma^\prime},
\label{hamiltonian}
\end{eqnarray*}
where $\hat{c}_{i,\sigma}$ denotes the annihilation operator of a fermion with spin $\sigma$, which takes $\mathcal{N}$ different values, at lattice site $i$, and $\hat{n}_{i,\sigma}\equiv\hat{c}_{i,\sigma}^\dagger\hat{c}_{i,\sigma}$ counts the local number of $\sigma$ fermions. Here, we take the strong-coupling limit ($U/t\gg 1$) of $\hat{\mathcal{H}}_{\rm Hub}$ under unit-filling conditions, with a particular emphasis on the spin degrees of freedom. This leads to the SU($\mathcal{N}$) Heisenberg model in the fundamental representation~\cite{Zhang2006,Beach2009,Hermele2009-fp}
\begin{eqnarray}
\hat{\mathcal{H}}_{\rm spin}&=&J\sum_{\langle i,j\rangle}\hat{\mathcal{S}}_{i,j}~~\left(J\equiv \frac{2t^2}{U}\right)
\label{hamiltonian2}
\end{eqnarray}
with $\hat{\mathcal{S}}_{i,j}\equiv |i,\sigma;j,\sigma^\prime\rangle\langle i,\sigma^\prime;j,\sigma|$, which swaps the spins at neighboring two sites. The swapping operator $\hat{\mathcal{S}}$ can be expressed as the linear combination of the bilinear terms of $N^2-1$ SU($\mathcal{N}$) generators with equal coefficients~\cite{Zhang2006,Beach2009}, guaranteeing that $\hat{\mathcal{H}}_{\rm spin}$ possesses the global SU($\mathcal{N}$) symmetry. 
Due to the SU($\mathcal{N}$) symmetry in the spin-swapping interactions, the global populations of each component, $N_\sigma \equiv \sum_i \hat{n}_{i,\sigma}$, become good quantum numbers. 
Below, we consider the case of two-dimensional square optical lattice with lattice constant $a$, which is relevant to many interesting SCES materials. Our main focus is on the $\mathcal{N}=6$ scenario, which represents the typical case of $^{173}$Yb with nuclear spin components $\sigma=\pm 5/2, \pm 3/2, \pm 1/2$, but the other cases including $\mathcal{N}=10$ for $^{87}{\rm Sr}$ are analogous. 

\begin{figure}[t]
\includegraphics[scale=0.7]{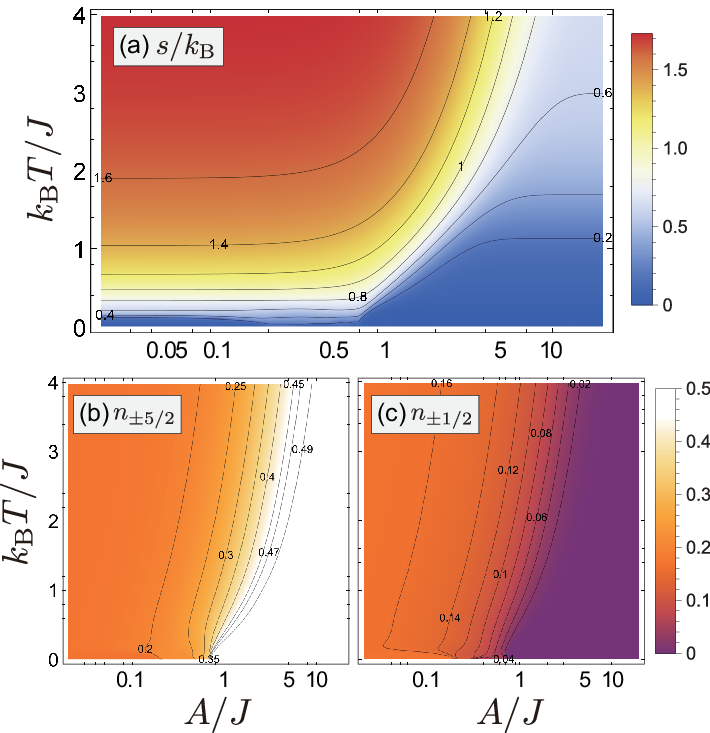}
\caption{\label{fig1}
(a) Entropy per site $s/k_{\rm B}$, (b) population rate of the spin components $\sigma={\pm 5/2}$, and (c) that of $\sigma={\pm 1/2}$ as a function of temperature $k_{\rm B}T/J$ and the strength of uniform quadratic Zeeman field $A/J$, obtained by the FTL method~\cite{Jakli1994,Prelovek2013} for an 18-site rhombic cluster. The population rate of the remaining components can be calculated with $n_{\pm 3/2}=0.5-n_{\pm 5/2}-n_{\pm 1/2}$. 
} 
\end{figure}
%
First, let us see the entropy characteristics of the spin Hamiltonian $\hat{\mathcal{H}}_{\rm spin}$ [Eq.~\eqref{hamiltonian2}] under the presence of uniform ``quadratic Zeeman-type'' field
\begin{eqnarray}
\hat{\mathcal{H}}_{\rm A}&=&-\frac{A}{2}\sum_i\left(\hat{S}^z_i\right)^2~~(A\geq 0),
\label{uni_field}
\end{eqnarray}
where $\hat{S}^z_i$ is the $z$ component of the spin-5/2 operator at site $i$. This field term plays the role of the chemical potentials for each spin component and introduces a population imbalance of the form $N_{\pm 5/2}>N_{\pm 3/2}>N_{\pm 1/2}$, where $N_{\pm\sigma}\equiv N_{\sigma}=N_{-\sigma}$.

To calculate the entropy of the system as a function of the temperature $T$ and the field strength $A$, we perform extensive numerical computations using the finite-temperature Lanczos (FTL) method~\cite{Jakli1994,Prelovek2013}. For imbalanced six-component mixtures, the complete Hilbert space is decomposed into the subspaces labeled by total numbers of atoms $N_\sigma$ of each spin component with $\sum_\sigma N_\sigma=N_{\rm site}$, where $N_{\rm site}$ is the number of lattice sites. We perform the FTL calculation on the 18-site $3\sqrt{2}\times 3\sqrt{2}$ rhombic cluster under periodic boundary conditions, for which the dimension of the largest subspace (with $N_\sigma=3$ for all $\sigma$) is given by 137,225,088,000. To improve the accuracy in the large-$A$ region, we carry out the full exact diagonalization for subspaces whose dimentions are less than 50,000. We confirm that the finite-size effect is sufficiently small for $T\gtrsim 0.3 J/k_{\rm B}$, by checking the convergence with the results for a 16-site cluster. In addition, as a reference  for comparison, we also calculate the entropy characteristics of the simple SU(2) Heisenberg model on square lattice for the 32-site $4\sqrt{2}\times 4\sqrt{2}$ cluster. 

In Figs~\ref{fig1}(a),~\ref{fig1}(b), and~\ref{fig1}(c), we show the entropy per site, $s/k_{\rm B}$, the population rate, $n_\sigma\equiv N_\sigma/N_{\rm site}$, for $\sigma={\pm 5/2}$, and that for $\sigma={\pm 1/2}$, respectively, as functions of the temperature $k_{\rm B}T/J$ and the field strength $A/J$. As can be seen in Fig.~\ref{fig1}(a), the entropy is larger for smaller $A$ at a given temperature. This is because when $A=0$ the six components are equally populated while only the two of six remain in the limit of $A\rightarrow \infty$ [see \ref{fig1}(b)]; the maximum entropy per site is given by $s^{({\rm max})}\approx 1.79 k_{\rm B}$ ($\approx 0.69 k_{\rm B}$) for six-component (two-component) systems. 

\begin{figure}[t]
\includegraphics[scale=0.64]{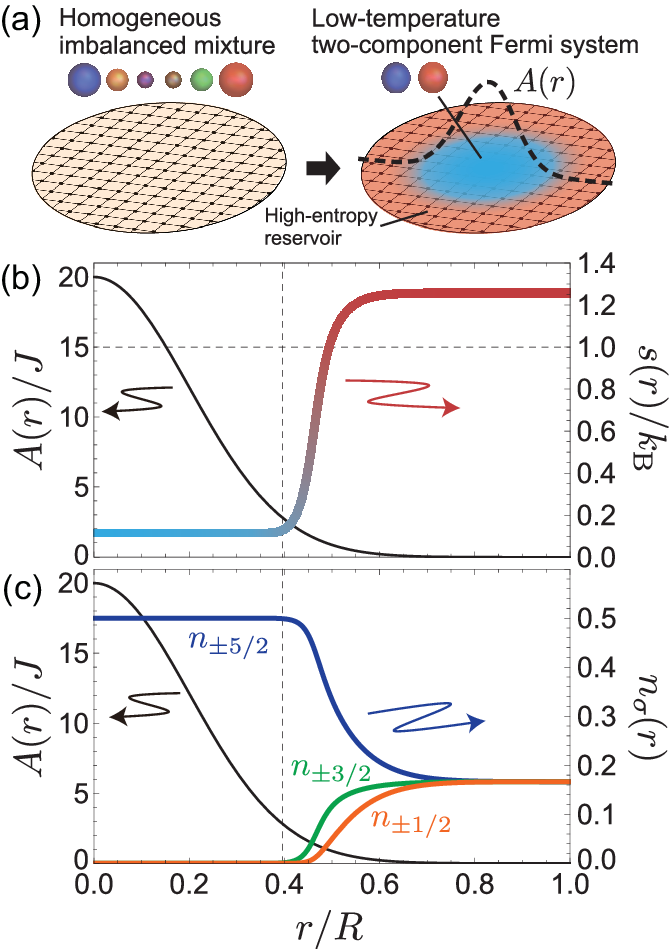}
\caption{\label{fig2}
(a) Sketch of the proposed cooling procedure. (b) and (c) Profiles of local entropy $s(r)/k_{\rm B}$ and population rate of each component $n_\sigma(r)$ after adiabatic insertion of the Gaussian field $A(r)$ with $A_0=20J$ and $w=0.2R$, respectively. The horizontal and vertical lines indicate the initial entropy per site $s_{\rm ini}/k_{\rm B}=1.0 k_{\rm B}$ and the radius of the SU(2) region $r_{\rm SU(2)}/R=0.396$, respectively. } 
\end{figure}

The field of the type described by Eq.~\eqref{uni_field} induces a population imbalance of the form $N_{\pm 5/2}>N_{\pm 3/2}>N_{\pm 1/2}$ as seen in Fig.~\ref{fig1}(b). Thus, by adiabatically introducing a similar field but with non-uniform intensity of the Gaussian shape (height $A_0\geq 0$; width $w\geq 0$), 
\begin{eqnarray} 
\hat{\mathcal{H}}_{\rm A}^\prime=-\sum_i \frac{A(r_i)}{2}\left(\hat{S}^z_i\right)^2~{\rm with}~A(r)=A_0e^{-\frac{r^2}{2w^2}},\label{nonuniform}
\end{eqnarray}
where $r$ is the distance from the center, into a homogeneous six-component mixture, it is expected that two of the six components, specifically $\sigma=\pm 5/2$, are selectively gathered to the central region of the entire system, resulting in the formation of a low-entropy pseudospin-1/2 subsystem with $|\pm 5/2\rangle\equiv|\uparrow\rangle,~|\downarrow\rangle$ surrounded by high-entropy reservoir of a six-component mixture as sketched in Fig.~\ref{fig2}(a).

To demonstrate the efficiency of this cooling procedure, let us consider the simple case where a sufficiently large number of sites exist inside a disk-shaped region of radius $R\gg a$, and treat the lattice coordinates as a continuous space. In addition, we employ the local density approximation (LDA)~\cite{Taie2022}, in which we assume that the local properties of the inhomogeneous system at position $r$ are given by the ones computed in a homogeneous system with field strength $A=A(r)$. Using the LDA, we can convert the data obtained by the FTL method for uniform fields (shown in Fig.~\ref{fig1}) into the distributions of the population $n_{\pm\sigma}(r)$ and of the local entropy $s(r)$ in the presence of the Gaussian field $A(r)$. Supposing that the initial entropy of a homogeneous six-component gas per site is $s_{\rm ini}$, we determine the temperature of the system $T$ after inserting the Gaussian field $A(r)$ (and the accompanying $s(r)$ and $n_{\pm\sigma}(r)$) such that the adiabatic condition $2\pi\int_0^R s(r)rdr/\pi R^2=s_{\rm ini}$ is satisfied.  

Figures~\ref{fig2}(b) and~\ref{fig2}(c) show the results for $s_{\rm ini}=1.0 k_{\rm B}$, $A_0=20J$ and $w=0.2R$. It can be seen that a large fraction of the entropy becomes stored in the surrounding six-component gas, as expected, along with the redistribution of the populations. As a result, the entropy per site at the center becomes much lower ($s(0)\approx 0.11k_{\rm B}$) than the initial value $s_{\rm ini}=1.0 k_{\rm B}$. Remarkably, the central region consisting only of two components has a sharp boundary at $r\approx r_{\rm SU(2)}$, which is defined by the condition $n_{\pm 5/2}\geq 0.499$, in spite of the smooth shape of the Gaussian field. This is attributed to the fact that the population of each component is a good quantum number due to the SU(6)-symmetric interactions. For the parameters of Figs.~\ref{fig2}(b) and~\ref{fig2}(c), the radius of the SU(2) region is $r_{\rm SU(2)}=0.396R$. The corresponding temperature becomes $T=0.753J/k_{\rm B}$.

In the experiments using $^{173}$Yb atoms, the field term described by Eq.~\eqref{nonuniform} can be realized using the light shifts by a linearly polarized light beam with a frequency detuned from the $^1S_0\leftrightarrow\, ^3P_1$ transition~\cite{Ozawa2018}. To achieve the population profile where the entropy reservoir subsystem consists of balanced six components, one needs to introduces a global population imbalance at the preparation stage of the initial homogeneous mixture. This is feasible by means of the optical-pumping technique~\cite{Taie2010,Ozawa2018}. The proper global population ratio is given by integrating $n_{\sigma}(r)$, and $N_{\pm 5/2}:N_{\pm 3/2}:N_{\pm 1/2}=0.258:0.126:0.116$ in the case shown in Fig.~\ref{fig2}. 

\begin{figure}[t]
\includegraphics[scale=0.55]{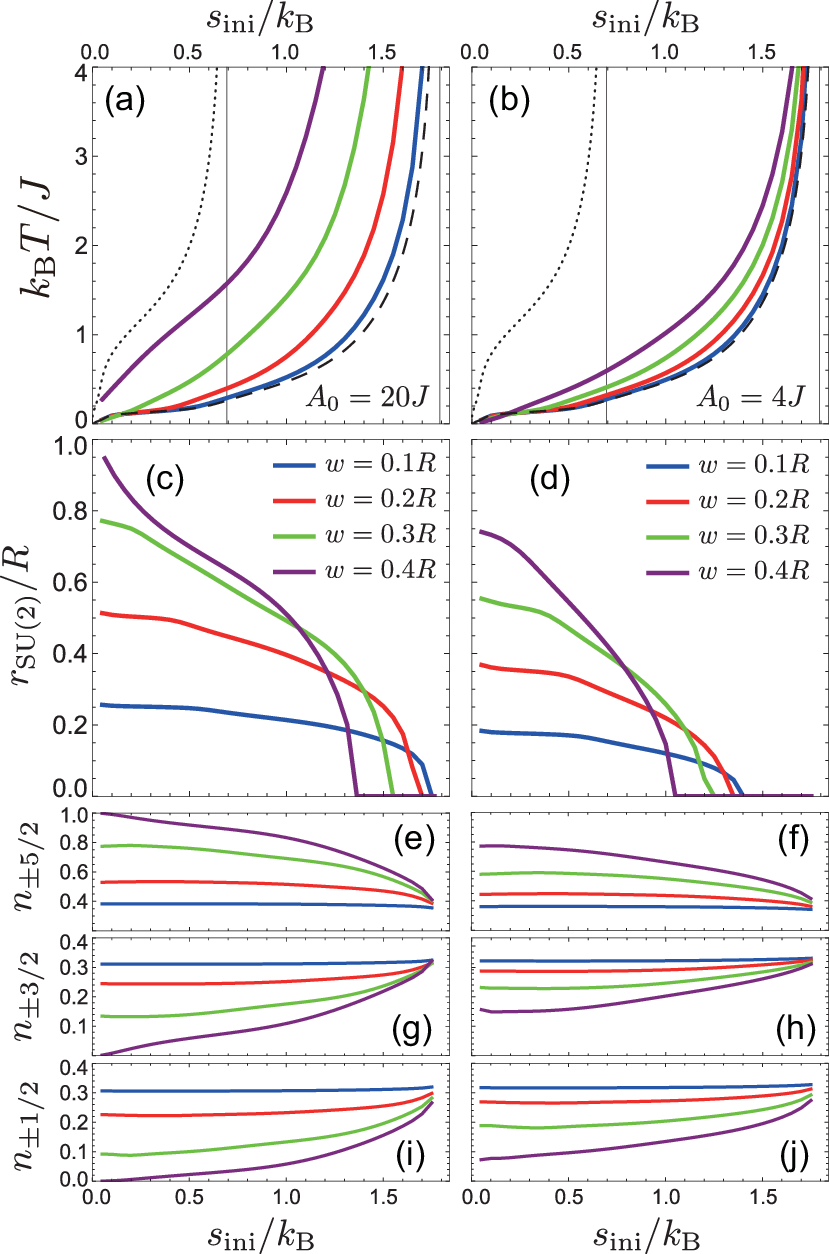}
\caption{\label{fig3}
(a,b) Temperatures and (c,d) radius of the SU(2) region, achieved by the proposed cooling procedure, as functions of the initial entropy per site. The ratio of the global population imbalance that has to be prepared in the initial homogeneous state is shown in (e-j). The height of the Gaussian field $A(r)$ are set to $A_0=20J$ for (a,c,e,g,i) and $A_0=4J$ for (b,d,f,h,j), and the results for different widths ($w=0.1R$-$0.4R$) are plotted together. The dotted [resp. dashed] curves in (a,b) show the temperature-entropy curve of a homogeneous SU(2) [resp. SU(6)] gas as a reference. The vertical lines represent the maximum entropy per site for two-component ($\approx 0.69 k_{\rm B}$) and for six-component ($\approx 1.79 k_{\rm B}$).} 
\end{figure}
In Figs.~\ref{fig3}(a-d), we present the cooling efficiency and the size of the central SU(2) region for various values of $w$. Panels (a,c) correspond to $A_0=20J$, while panels (b,d) correspond to $A_0=4J$. These results provide guidance on the required initial entropy of the mixed gas to attain the desired temperature and the size of the SU(2) region. To engineer a two-component Fermi system of radius $r_{\rm SU(2)}\gtrsim 0.4 R$ at temperature, say, $T= 0.5 J/k_{\rm B}$, 
the cooling curve indicates that one needs to prepare the initial six-component mixture with $s_{\rm ini} \approx 0.8 k_{\rm B}$ for $A_0=20J$ and $w=0.2R$, while the required entropy per site to achieve the same temperature is quite small ($\approx 0.05 k_{\rm B}$) if one uses a homogeneous two-component gas. The initial setup of the homogeneous mixture requires the suitable global population imbalance shown in Figs.~\ref{fig3}(e) and~\ref{fig3}(f), depending on the parameters. 

It can be seen from the comparison of the curves for different values of $w$ that the achievable temperature is lower for a tighter field potential in exchange for a smaller region of two-component Fermi atoms, as naturally expected. When the height of the Gaussian field is reduced, the comparison between Figs.~\ref{fig3}(a) and~\ref{fig3}(b) tells us that the cooling efficiency gets better while a lower initial entropy is required to prepare a large enough SU(2) region. Hence, the optimal setting for $A_0$ and $w$ is determined comprehensively by the achievable entropy of the initial homogeneous mixture, the target temperature, and the intended size of the SCES quantum simulator.

In summary, we have explored an entropy engineering scheme for two-component Fermi systems employing a multi-component mixture of atomic gases. This scheme involves the adiabatic insertion of a nonuniform field of the quadratic-Zeeman-type, which divides the system into a central low-entropy region with only two specific components and a surrounding $\mathcal{N}$-component entropy reservoir. Taking the case of a two-dimensional optical-lattice system of $^{173}{\rm Yb}$ atoms, which have $\mathcal{N}=6$ nuclear components with fully symmetric interactions in the ground state, we have presented the estimation of the cooling efficiency of this entropy engineering scheme.

In the experiment of Ref.~\onlinecite{Mazurenko2017}, which utilized the cooling method relying on the high motional degrees of freedom of a metallic state serving as an entropy reservoir, the lowest temperature achieved was estimated to be $T/t=0.25(2)/k_{\rm B}$ for a system of two-component fermions in a two-dimensional optical lattice, described by the Hubbard model with $U/t=7.2(2)$. This corresponds to $T\approx 0.9 J/k_{\rm B}$ in our energy unit $J\equiv 2t^2/U$. To attain the same temperature using the cooling method discussed here with $^{173}{\rm Yb}$ atoms, it is necessary to prepare a six-component mixture with initial entropy of $s_{\rm ini}= 1.08 k_{\rm B}$ in the spin part, considering a typical case of $A_0=20J$ and $w=0.2R$, according to Fig.~\ref{fig3}(a). While our focus has been on the unit-filling region of the entire system in this study, there exists a lower-density region in the metallic phase further outside ($r> R$) in an actual experimental situation. Therefore, these two methods could be used in conjunction, offering the expectation of achieving low enough temperatures for studying highly quantum phenomena in SCESs.

It is worth noting that the current cooling method is expected to be even more effective for a larger value of ${\mathcal{N}}$, including ${\mathcal{N}=10}$ for $^{87}{\rm Sr}$~\cite{DeSalvo2010,Tey2010,Zhang2014}, and can also be applied to multi-component systems without perfect SU($\mathcal{N}$) symmetry, although precise engineering of the shape of the spin-dependent field is necessary to achieve a sharp boundary of the SU(2) region. Furthermore, the method can be extended to quantum simulations of low-entropy states in SU($\mathcal{M}$) systems where $2<\mathcal{M}<\mathcal{N}$ by using a field that can selectively gather $\mathcal{M}$ out of $\mathcal{N}$ components in the central subsystem. This opens up possibilities for realizing exotic SU(3)~\cite{Tth2010,Bauer2012,Yamamoto2020,Motegi2023} and SU(4)~\cite{Corboz2011,Miyazaki2022} magnetism, which is also relevant to the physics of solid-state materials, including nematic liquid crystals~\cite{Nakatsuji2005,Tsunetsugu2006,Bhattacharjee2006}, transition metal oxides~\cite{Tokura2000}, and graphene~\cite{Goerbig2011}.

\begin{acknowledgments}
We would like to thank Y. Takahashi, Y. Takasu, S. Taie, and I. Danshita for useful discussions. The work of D.Y. was supported by JSPS KAKENHI Grant Nos.~21H05185, 22H01171, 23H01133, and JST PRESTO Grant No.~JPMJPR2118. The computational results presented were obtained in part
using the OCTOPUS at the Cybermedia Center, Osaka University. D.Y. and K.M. contributed equally to this work.
\end{acknowledgments}

\bibliography{apsbib.bib}

\end{document}